\documentclass{aa}
\usepackage{graphicx}
\usepackage{txfonts}
\usepackage{float}
\begin{document}

   \title{Spatially resolved spectroscopy across stellar surfaces. I.}

   \subtitle{Using exoplanet transits to analyze 3-D stellar atmospheres}

   \author{Dainis Dravins
          \inst{1},
               Hans-G\"{u}nter Ludwig
           \inst{2},
          Erik Dahl\'{e}n,
        \inst{1}
          \and
        Hiva Pazira
          \inst{1,3} \fnmsep 
                }
%
%
   \institute{Lund Observatory, Box 43, SE-22100 Lund, Sweden\\
              \email{dainis@astro.lu.se}
\and
     Zentrum f\"{u}r Astronomie der Universit\"{a}t Heidelberg, Landessternwarte K\"{o}nigstuhl, DE--69117 Heidelberg, Germany\\
              \email{hludwig@lsw.uni-heidelberg.de}
\and 
                Present address: Department of Astronomy, AlbaNova University Center, SE--10691 Stockholm, Sweden\\
             }

   \titlerunning{Spatially resolved stellar spectroscopy. I.}
   \authorrunning{D.\ Dravins et al.} 
 
   \date{Received XXX YY, 2017; accepted XXX YY, 2017}

   \abstract
   {High-precision stellar analyses require hydrodynamic modeling to interpret chemical abundances or oscillation modes.  Exoplanet atmosphere studies require stellar background spectra to be known along the transit path while detection of Earth analogs require stellar microvariability to be understood.  Hydrodynamic 3-D models can be computed for widely different stars but have been tested in detail only for the Sun with its resolved surface features.  Model predictions include spectral line shapes, asymmetries, and wavelength shifts, and their center-to-limb changes across stellar disks.}
   {To observe high-resolution spectral line profiles across spatially highly resolved stellar surfaces, which are free from the effects of spatial smearing and rotational broadening present in full-disk spectra, enabling comparisons to synthetic profiles from 3-D models.}
   {During exoplanet transits, successive stellar surface portions become hidden and differential spectroscopy between various transit phases provides spectra of small surface segments temporarily hidden behind the planet.  Planets cover no more than $\sim$\,1\% of any main-sequence star, enabling high spatial resolution but demanding very precise observations.  Realistically measurable quantities are identified through simulated observations of synthetic spectral lines. }
   {In normal stars, line profile ratios between various transit phases may vary by $\sim$0.5\,\%, requiring S/N ratios $\gtrsim$\,5,000 for meaningful spectral reconstruction.  While not yet realistic for individual spectral lines, this is achievable for cool stars by averaging over numerous lines with similar parameters.}
   {For bright host stars of large transiting planets, spatially resolved spectroscopy is currently practical.  More observable targets are likely to be found in the near future by ongoing photometric searches. }


   \maketitle

\section{Precision stellar and exoplanet studies}

Three-dimensional and time-dependent hydrodynamic simulations are now established as realistic descriptions for the surfaces and convective photospheres of various classes of stars and must be used in any high-precision determinations of stellar properties.  This applies to determinations of stellar abundances, perhaps ultimately aiming at accuracies approaching $\pm$0.01 dex, which would permit the study of different chemistry in stars with or without specific types of planets or the application of precise chemical tagging in searches for stars that once formed together.  This also applies to accurate analyses of stellar oscillations in which exact dimensions and depth structures of the photospheres must be obtained from models of the convective near-surface layers. This is required for stellar differential rotation and stellar gas dynamics, which are deduced from subtle signatures of line profiles, as are magnetic fields. 

Atmospheric properties of exoplanets are deduced differentially to the stellar spectrum, which thus must be precisely known.  For transiting planets, this requires a knowledge of the varying stellar spectra at the positions along the transit path of the planet; these spectra are filtered through the exoplanetary atmosphere rather than the spectrum of the stellar disk-integrated flux.  A challenge is to find and study `true' Earth analogs whose sizes and orbits are comparable to the terrestrial values.  Since the miniscule effects in induced radial velocity of the host star and the tiny photometric amplitudes during any transit are much smaller than physical stellar variability, these effects must somehow be calibrated and corrected for.  One approach might be to correlate spectral line variations to those in photometric brightness, measuring the latter to calibrate the former.

All such precision studies of stars and exoplanets require an understanding of 3-D and time-variable stellar hydrodynamics, as well as spectral line formation in such dynamic atmospheres and how the ensuing properties change across stellar surfaces.  Hydrodynamic models can now be produced for widely different stars, ranging from white dwarfs to supergiants, and with all sorts of metallicities \citep{beecketal12, freytagetal12, magicetal13, tremblayetal13}; Fig.\ 1.  Although, in principle, such simulations do not have freely tunable physical parameters, their complexity implies that they must apply various physical, mathematical, and numerical approximations to be manageable.  The uncertainties increase for stellar types that are increasingly deviant from the Sun.  These uncertainties, for example, include the largest spatial scales on which global oscillations should be modeled in non-solar type stars, and whether there are phenomena corresponding to solar supergranulation.  Such models well reproduce the details of solar spectral line profiles and the fine structure (granulation) observed across the solar surface, its time evolution, and its interaction with magnetic fields.  However, the verification (or falsification) of such models is more challenging for other stars, where the direct observation of highly resolved stellar surfaces remains unrealistic.

\begin{figure}[H]
\centering
\includegraphics[width=\hsize]{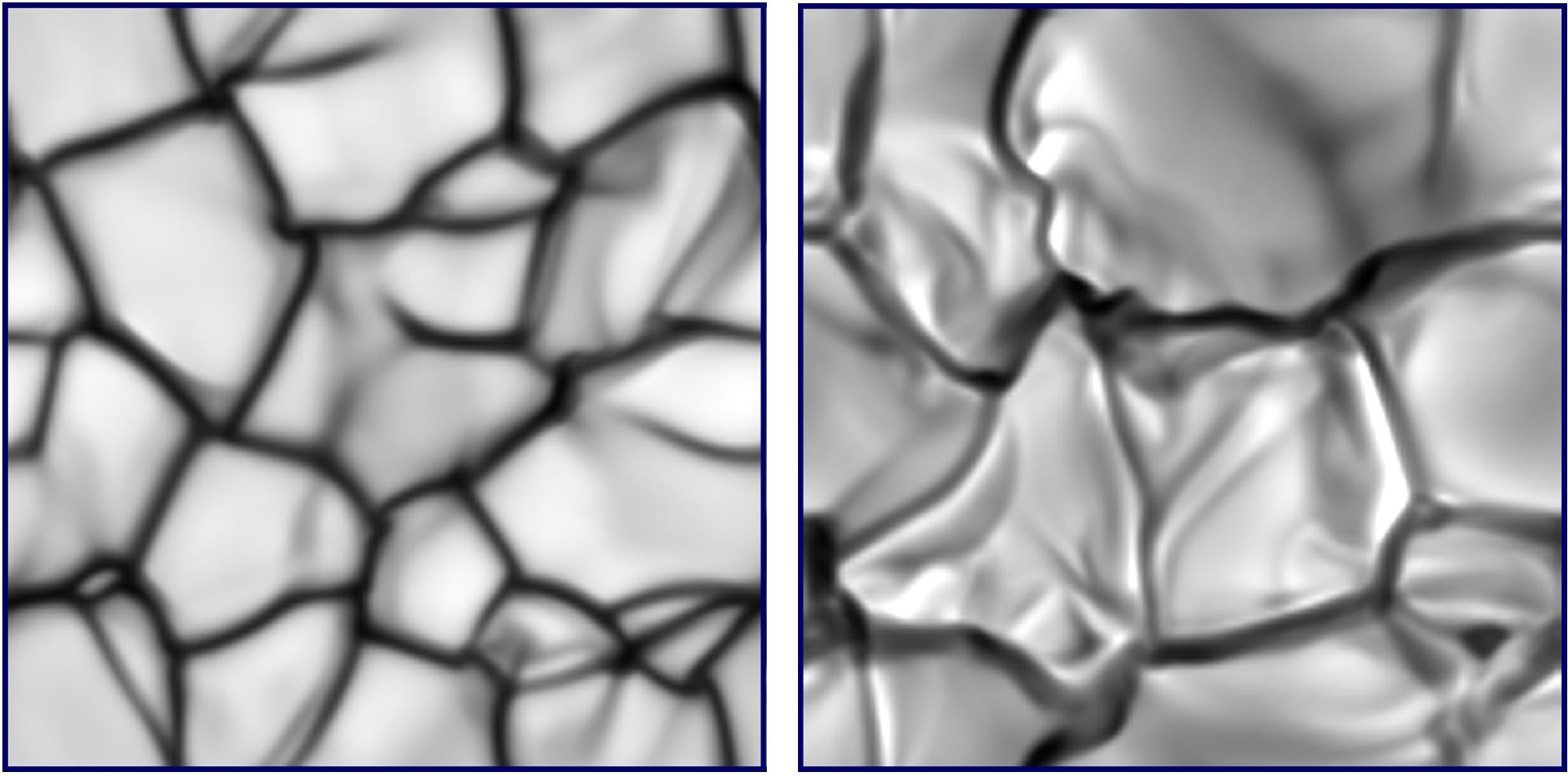}
\caption{Simulations of 3-D hydrodynamics in stellar atmospheres. Examples of emergent intensity patterns during granular evolution on the surface of a 12,000 K white dwarf (left) and a 3,800 K red giant, obtained from modeling with the CO\,$^5$BOLD code.  The areas differ by many orders of magnitude: 7$\times$7~km$^{2}$ for the white dwarf and 23$\times$23~R$_{\odot}$$^{2}$ for the giant. }
\label{fig:3-d_modeling}
\end{figure}

Possibilities for dissecting fine structure on stellar surfaces are opening up through the exploitation of exoplanets as probes, scanning across the stellar surface.  During a transit, an exoplanet covers successive segments of the stellar disk and differential spectroscopy between each transit phase and outside transit can provide spectra of each particular surface segment that was temporarily hidden behind the planet.  That is the topic of this project, where the present Paper I surveys diagnostic tools and tests that are available for examining hydrodynamic model atmospheres for different classes of main-sequence stars.  In particular, this survey uses synthetic photospheric line profiles for differently strong lines, and of varying excitation potential, from a series of stellar models to predict observable spectral features and to identify those that appear most realistic to measure with current or near-future facilities.  Starting with Paper II \citep{dravinsetal17}, we apply these methods to stellar observations to reconstruct spectral lines across the disks of a few among the best-observed transiting systems, identifying spectroscopic features specific to atmospheric hydrodynamics.

\section{Hydrodynamic stellar atmospheres}

Several groups have developed simulations of the 3-D radiative hydrodynamics of convective stellar photospheres.  These normally are of the type `box-in-a-star', i.e., a simulation volume of limited spatial extent is immersed into the stellar surface layers, with an energy flux entering from below the volume and escaping as radiation from its top.  The resulting atmospheric structure depends on parameters such as the effective temperature, surface gravity, opacity of the gas (i.e., stellar metallicity), vertical and horizontal scales of the simulation volume and its boundary conditions, the types of waves permitted (oscillations over which scales; perhaps also shock waves), temporal and spatial step sizes, and the duration of the simulation toward statistical stability.

Notable model families, from which simulations for stars of different spectral types have been computed include the CO\,$^5$BOLD grid \citep{freytagetal12, ludwigkucinskas12, tremblayetal15, wedemeyeretal13} and the StaggerGrid \citep{magicetal13, magicasplund14}.  Evaluations of different models are given by \citet{pereiraetal13} and \citet{trampedachetal13}.

\subsection{Full stellar-disk diagnostics}

Some supergiant stars can already be spatially resolved by the largest telescopes and several more by optical interferometers.  Although full 2-D imaging of stellar surfaces still remains a challenge \citep{patruetal10}, effects from 3-D atmospheric structure are revealed by surface brightness variations \citep{chiavassabigot14, chiavassaetal09, chiavassaetal10a, chiavassaetal10b, chiavassaetal11b, creechetal10}.  Even nearby main-sequence stars start to get spatially resolved by such interferometric studies, e.g., Procyon \citep{chiavassaetal12} or $\alpha$\,Cen \citep{kervellaetal17}, potentially even revealing signatures of exoplanet transits \citep{chiavassaetal14}.   Limb-darkening functions differ between 1-D and 3\mbox{-}D models, depend on the amounts of scattering and absorption, and are testable with exoplanet transit photometry \citep{espinozajordan15, espinozajordan16, hayeketal12, magicetal15}.  

Modeled evolution of surface inhomogeneities provide a number of testable predictions, such as photometric or radial-velocity flickering of integrated starlight.  Stellar brightness varies in a somewhat random fashion, as differently bright granular structures develop across the surface \citep{cranmeretal14, ludwig06}.  The effects are greater for giant stars, where the small-number statistics of fewer granules across the star induces a greater variability, as observed \citep{bastienetal13, bastienetal16, mathuretal11} and as modeled \citep{samadietal13a, samadietal13b}. For the Sun, with $\sim$10$^6$ granules across its surface, each with a $\sim$20$\%$ intensity contrast, the amplitude of variability in integrated sunlight (assuming random variability among the granules) can be expected to be on the order of this contrast divided by the square root of the number of granules, i.e., $\sim$2\,10$^{-4}$, measurable with photometric instruments in space.  Since the contrast is greater at short optical wavelengths (as expected for a given temperature contrast in the photosphere), also the amplitude of flickering is greater in the blue than in the red.  In specific wavelength regions comprising numerous temperature-sensitive spectral lines, such as the G-band, the photometric  flickering may be expected to be greater but does not yet appear to have been either computed nor observed. 

An analogous variability in the disk-averaged radial velocity on a level of some m\,s$^{-1}$ is a concern in radial-velocity searches for low-mass exoplanets and for the confirmation of exoplanet candidates found from transits \citep{ceglaetal13, ceglaetal14, dumusqueetal11a}.  Again, the amplitude of flickering can be expected to increase at shorter wavelengths, where the granular contrast is greater.  Further, the astrometric location of the stellar photocenter may wander and is detectable with space astrometry \citep{chiavassaetal11a, pasquatoetal11}.  

Other effects include interpretation of stellar p-mode oscillations, where 3-D models have to be applied for any more precise representation of stellar near-surface layers, detecting inconsistencies that may indicate which modeling parameters need adjustment \citep{baldnerschou12, kallingeretal14, piauetal14, schou15}.

\begin{figure}[H]
\centering
\includegraphics[width=\hsize]{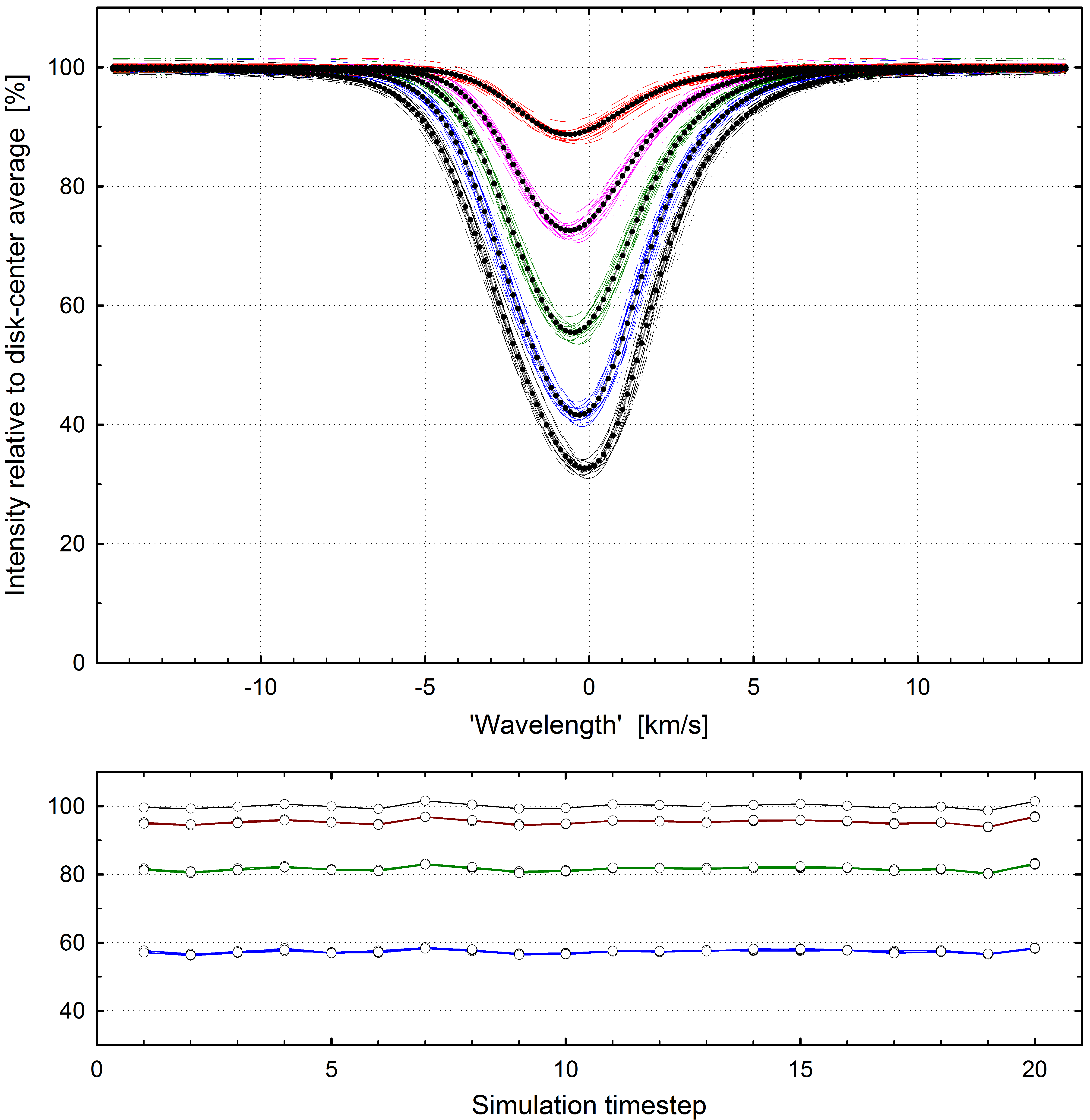}
\caption{Top: Build-up of synthetic Fe~I lines at stellar disk center ($\mu$ = cos$~\theta$ = 1.0) from a CO\,$^5$BOLD model of a dwarf star with approximate spectral type F7~V (T$_{\textrm{eff}}$ = 6250 K, log~$\varg$ [cgs] = 4.5).  Thin colored lines are spatial averages over the simulation area for each of 20 temporal snapshots; bold curves are the total averages.  All five different line strengths are for $\lambda$ = 620 nm, $\chi$ = 3 eV.  The convection patterns especially cause the weaker lines to become asymmetric and wavelength-shifted.  Bottom: Evolution of the continuum intensity during the simulation sequence for four different center-to-limb positions $\mu$ = 1, 0.87, 0.59, 0.21.  For off-center positions, slight differences exist among the four different azimuth angles ($\psi$=0, $\pi$/2, $\pi$, 3$\pi$/2 rad).}
\label{fig:line_buildup}
\end{figure}

\subsection{Spatially resolved diagnostics}

While such full-disk measures reveal the presence of inhomogeneities on stellar surfaces, they do not provide particularly sharp diagnostic tools for more detailed model simulations.  In order to falsify and successively refine 3-D modeling, quantities have to be identified that are both realistically observable and display nontrivial properties.  Solar models can be compared to solar surface structure, but we seek to discern which observable phenomena can be predicted from 3-D simulations for other stars.

In the absence of detailed surface imaging, the most detailed tests may come from analyses of spectral line profiles.  Using the output from hydrodynamic simulations as spatially varying model atmospheres, synthetic spectral line profiles can be computed as temporal and spatial averages over the simulation sequences \citep{beecketal13,holzreutersolanki13, pereiraetal13}. Details of the atmospheric structure and dynamics are reflected in the exact profiles of photospheric absorption lines and in their center-to-limb variations. These changes are gradual and not sensitive to the exact spatial resolution across the star.  As opposed to classical and stationary atmospheres, lines in general become asymmetric and shifted in wavelength \citep[e.g.][]{asplundetal00}; this is caused by the statistical bias of more numerous blueshifted photons from hot and rising surface elements (Figs.\ \ref{fig:3-d_modeling}-\ref{fig:line_buildup}).  The amount (and even sign) of the effect differs among lines of different strength, excitation potential, ionization level, and wavelength region.  High-temperature regions are better diagnosed by lines from ionized species, low-temperature regions by molecular lines.  For F-, G-, and K-type main-sequence stars, convective blueshifts increase with increasing stellar temperature and increasing luminosity.  Center-to-limb changes depend on the relative amplitudes of horizontal and vertical velocities and may differ between stars with `smooth' or `corrugated’ surfaces; these changes further depend on whether line formation is computed in one or in multiple dimensions and whether or not local thermodynamic equilibrium is assumed.

\begin{figure}[H]
\centering
\includegraphics[width=\hsize]{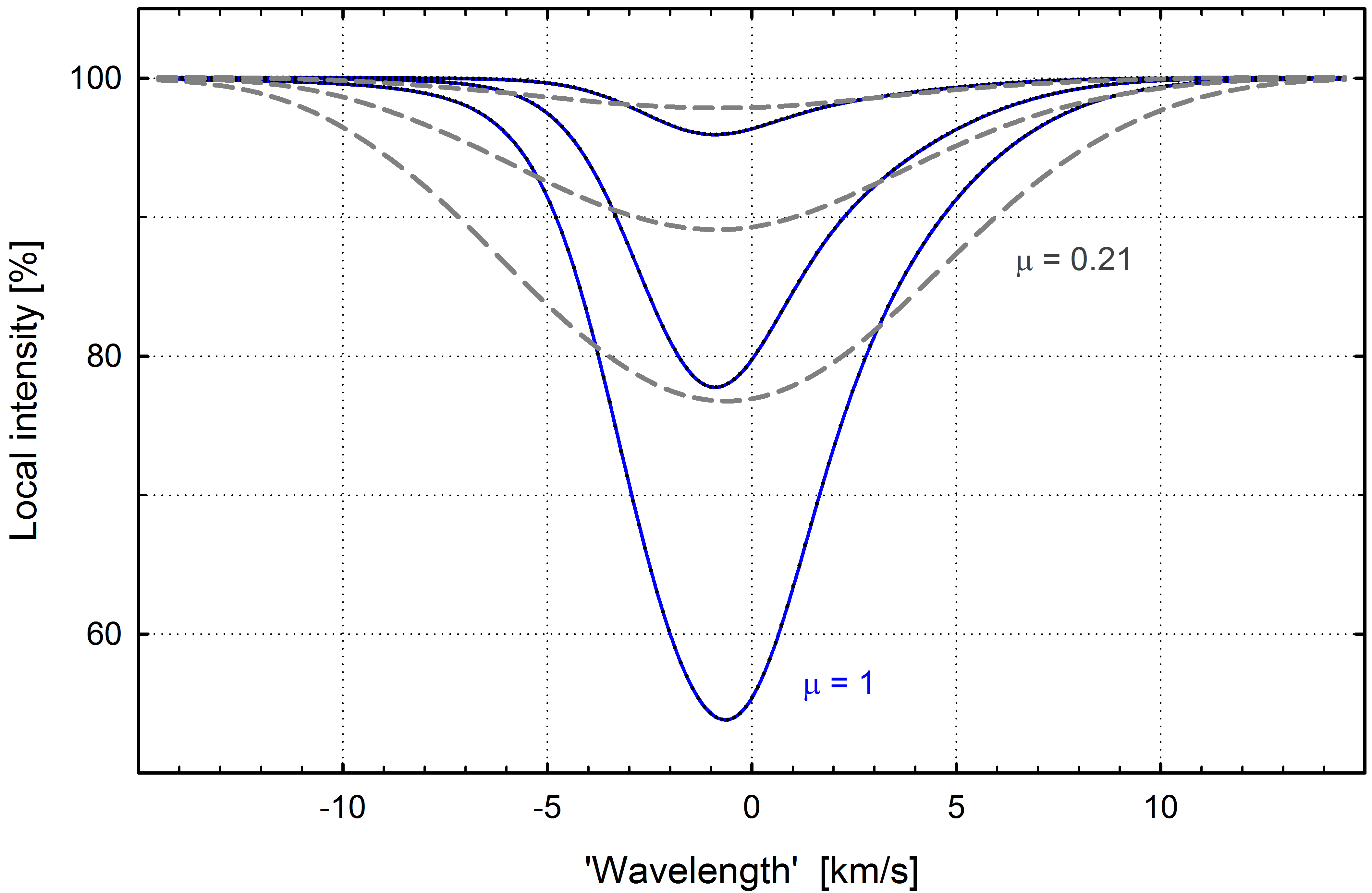}
\caption{Spectral line changes across a stellar disk, predicted from 3\mbox{-}D simulations with the CO\,$^5$BOLD code of an `F3~V' main-sequence star.  Fe I lines of three different strengths ($\lambda$ = 620 nm, $\chi$ = 3 eV) are shown at disk center (viewing angle against the normal to the stellar surface $\theta$ = 0; cos\,$\theta$ = $\mu$ = 1) and close to the limb ($\mu$ = 0.21)   Relative to a frame of rest, lines are displaced toward shorter wavelengths (`convective blueshift'), which is an effect that changes from disk center toward the limb.  Profiles are here normalized to the local continuum intensity at each center-to-limb position; the wavelength scale is in equivalent Doppler velocities.}
\label{fig:linewidths}
\end{figure}

When spatially resolved across the Sun, the amount of convective line shift can be seen to decrease from disk center toward the limb.  This solar `limb effect' is well understood in terms of 3-D hydrodynamics with the hottest and brightest elements predominantly moving upward.  This correlation between Doppler-shifting vertical velocities along the lines of sight produces a convective blueshift near disk center but the lack of such a correlation for horizontal velocities does not result in any significant wavelength shifts near the limb.  There, however, the absorption lines tend to become broader, reflecting the greater amplitudes of horizontal velocities (which are then the Doppler-broadening velocities along the line of sight) as a consequence of granular convective flows turning over in a photosphere whose vertical scale height is smaller than the lateral extent of the granules (Fig.\ \ref{fig:linewidths}).  Studies of the solar spectrum under more stellar-like conditions  include signatures in integrated sunlight during a transit of Venus \citep{chiavassaetal15} and during a partial solar eclipse \citep{reinersetal16a}, of convective line shifts and line asymmetries \citep{delacruzetal11, dravins08, molaromonai12, reinersetal16b}, center-to-limb changes of line profile shapes \citep{koesterkeetal08}, and temporal variability of line profiles and bisectors in disk-integrated sunlight \citep{dumusqueetal15, marchwinskietal15}.

Center-to-limb changes largely depend on the relative amplitudes of horizontal and vertical velocities, whose Doppler shifts contribute differently to the line broadening.  Such effects may vary among different types of stars.  Stars, where the geometrical surface of a given optical depth is `corrugated'  with `hills' and `valleys', should instead show an increased blueshift toward the limb.  This is predicted in models for the F-star Procyon of T{$\rm_{eff}$}= 6600~K \citep{allendeetal02, dravinsnordlund90a, dravinsnordlund90b}.  Although the velocities on the star are horizontally symmetric, near the limb one predominantly views the horizontal flows on the slopes of the hills facing the observer.  These approaching flows appear blueshifted while the equivalent redshifted components remain invisible behind these hills and an enhanced blueshift results.  The depth dependence of atmospheric properties causes such effects to depend on the oscillator strength, excitation potential, ionization level, and wavelength region of the line.  Obviously, the availability of such spectral data could open up stellar surface structure to detailed observational study \citep{dravinsnordlund90b, dravinsetal05}. 

The exact line shapes and shifts offer non-LTE diagnostics.  Even if modeling in local thermodynamic equilibrium would predict similar line shapes, the convective wavelength shifts (and their center-to-limb changes) could still be different.  For lines from neutral metals in solar-type stars, non-LTE calculations produce smaller convective blueshifts because the fastest upflows occur in the hottest granular elements.  These emit strong ultraviolet flux, ionizing the gas above, thus decreasing the most blueshifted line absorption contributions from neutral species and lessening their contribution \citep{amarsietal16b, dravinsnordlund90a, shchukinaetal05}.  Further effects enter when not only the atmospheric modeling but also the line formation is treated in three dimensions \citep{holzreutersolanki13}.  

Line formation details are also reflected in their time variability.  One observes an increased level of temporal spectral line variability already
on the Sun,  when observing an area of given size at different locations further away from the disk center.  This can be understood since the horizontal velocities are greater, and when observed under an inclined line of sight, have a larger variance \citep{dravinsnordlund90a}.  On a corrugated star, observed near its limb, a further element of variability enters since the swaying and corrugated stellar surface sometimes hides some granules from direct view. 

Some of these effects can be seen in integrated starlight, although their observability is constrained by the averaging of different intrinsic line profiles from different parts across the stellar disk, broadened by (possibly differential) stellar rotation.  Convective line shifts and line asymmetries have been analyzed for a number of stars and stellar models by \citet{allendeetal13, basturketal11, chiavassa11c, dravins08, dravinsetal05, pasquinietal11, ramirezetal08, ramirezetal09}. The effects on line formation and deduced chemical abundances have been analyzed by \citet{amarsietal16a, asplund05,  bergemann14, bergemannetal16, gallagheretal17, grevesseetal15, hayeketaL10, holzreutersolanki12, holzreutersolanki13, holzreutersolanki15, klevasetal16, lindetal13, lindetal17, magicetal14, scottetal15a, scottetal15b, steffenetal12, thygesenetal17}. Further studies exist for various classes of giant and supergiant stars, however, these are not discussed further here.

\begin{figure}[H]
\centering
\includegraphics[width=\hsize]{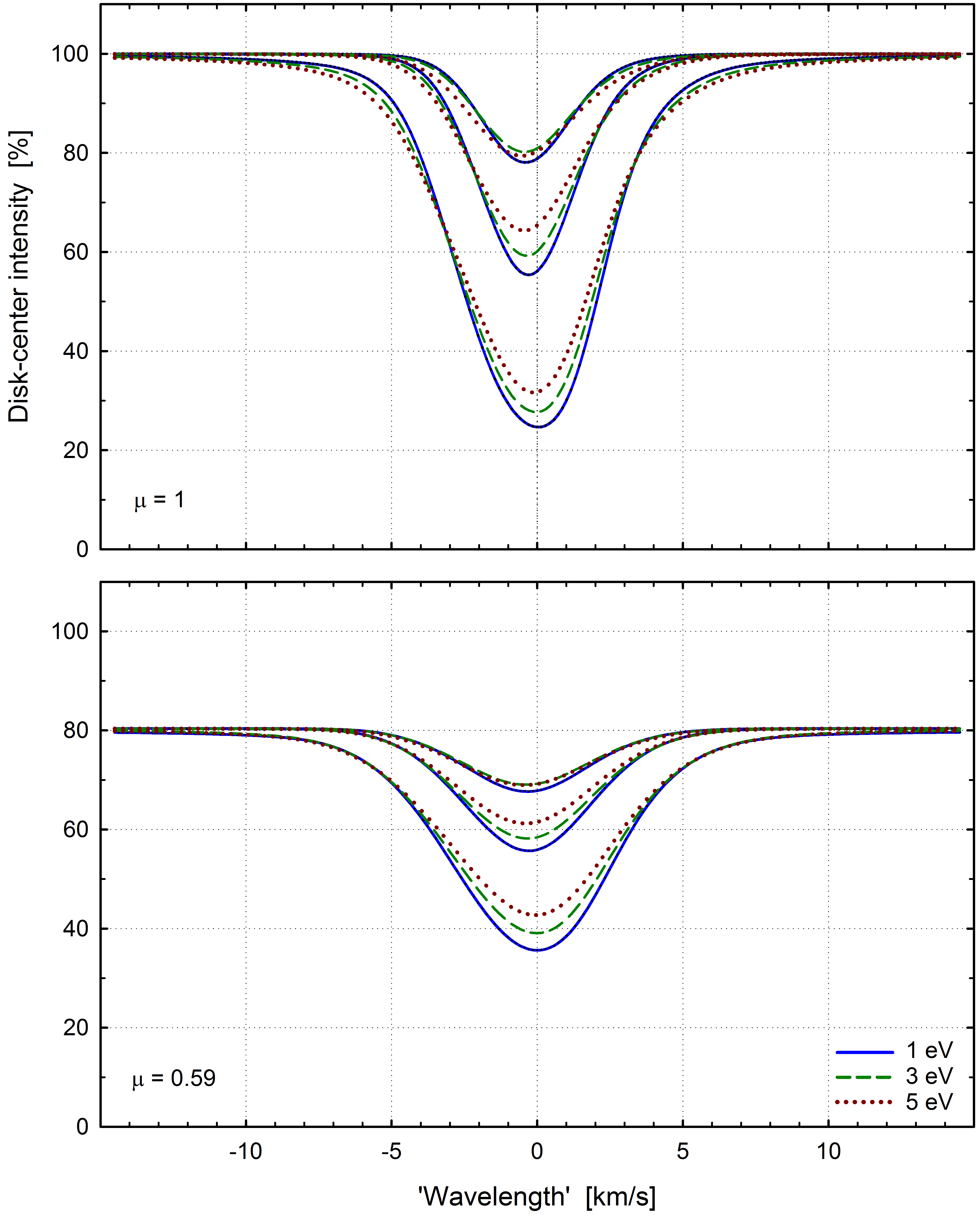}
\caption{Spectral line properties also depend on the excitation potential: Synthetic  Fe~I lines for a `Solar G2~V' model.  Top: Plot of disk center ($\mu$ = cos$\,\theta$ = 1.0), three line strengths, and three excitation potentials $\chi$ = 1, 3, and 5 eV is shown.  Bottom: Same for disk position $\mu$ = cos$\,\theta$ = 0.59.  The decreased intensity reflects the limb darkening at $\lambda$ = 620 nm.  Higher excitation lines tend to be shallower and broader although the exact properties depend on the detailed line formation for each transition.}
\label{fig:excitationpot}
\end{figure}

\subsection{Exoplanet detection limits}

Detailed knowledge of stellar spectra is required not only to study stellar atmospheres per se, but also to observe subtle differences in these spectra that may be caused by exoplanets.  One observes the spectrum during a transit, when some background starlight is seeping through the atmosphere of the exoplanet.  By comparing this spectrum to that without a planet, chemical signatures in the exoplanet atmosphere may be identified.  This spectrum of background starlight, however, is not of the flux integrated over the full stellar disk, but that of a location behind the planet. Thus, the need is to know the varying stellar signal along the exoplanet transit path.

A major challenge is to find `truly' Earth-like exoplanets, i.e., such planets with sizes and orbits comparable to those of the Earth around stars similar to the Sun \citep{fischeretal16}.  However, the diminutive signals induced by such a planet in either stellar radial-velocity variations or in photometric transit amplitudes are very much smaller for solar-type stars than for stellar intrinsic variability, demanding a detailed understanding of the latter \citep{borgnietetal15, ceglaetal12, ceglaetal13, dumusqueetal11a, dumusqueetal11b, herreroetal16, korhonenetal15, lagrangeetal10, lanzaetal11, lovisetlal11, meunierlagrange13, meunieretal10}. 

With $\sim$10$^6$ granules across the surface of a solar-type star, each with a velocity amplitude of  $\sim$2 km\,s$^{-1}$ (assuming these to evolve at random), the radial-velocity fluctuations in integrated starlight can be expected to be on order of this number, divided by the square root of  10$^6$, i.e., $\sim$2 m\,s$^{-1}$, close to what actually is observed.  A temporary radial-velocity excursion results from an increase of the number and/or contrast of bright and rising granules.  Given the correlation between granular temperature (brightness) and their rising motion, such excursions of convective blueshift must correlate with those in brightness.  Irradiance fluctuations in integrated starlight are measurable with photometry from space.   If spectral line variations can be calibrated against those in photometric brightness, one could measure the latter to correct the former to its average value.  Theoretical calibrations should become possible from 3-D models, perhaps separately in different wavelength regions and in different spectral features, but the model reliability must of course be verified.

A further issue is the presence of stellar magnetic structures such as spots.  Some (probably most) stars have some (magnetic or thermal) dark or bright spots on their surfaces and transiting exoplanets may happen to cross in front of some of those spots.  When this happens, the (probably magnetic) spectral signature from the particular spot area is temporarily hidden and could -- if sufficient signal to noise can be achieved -- be retrieved as the difference from the signal outside transit.  A spot transit can be identified in the photometric signature and several such instances have already been observed.  The varying amplitudes in different photometric colors may yield an indication of the spot temperature while a retrieved spectrum (with possibly magnetic Zeeman signatures) could provide quite detailed information on stellar spots and atmospheric magnetohydrodynamics.

Besides limb darkening, continuum radiation also carries some small degree of polarization; this depends on the wavelength and could be detected in exoplanet transit polarimetry \citep{kostogryzetal15, kostogryzetal16, wiktorowiczlaughlin14}.

\begin{figure*}
\sidecaption
  \includegraphics[width=12cm]{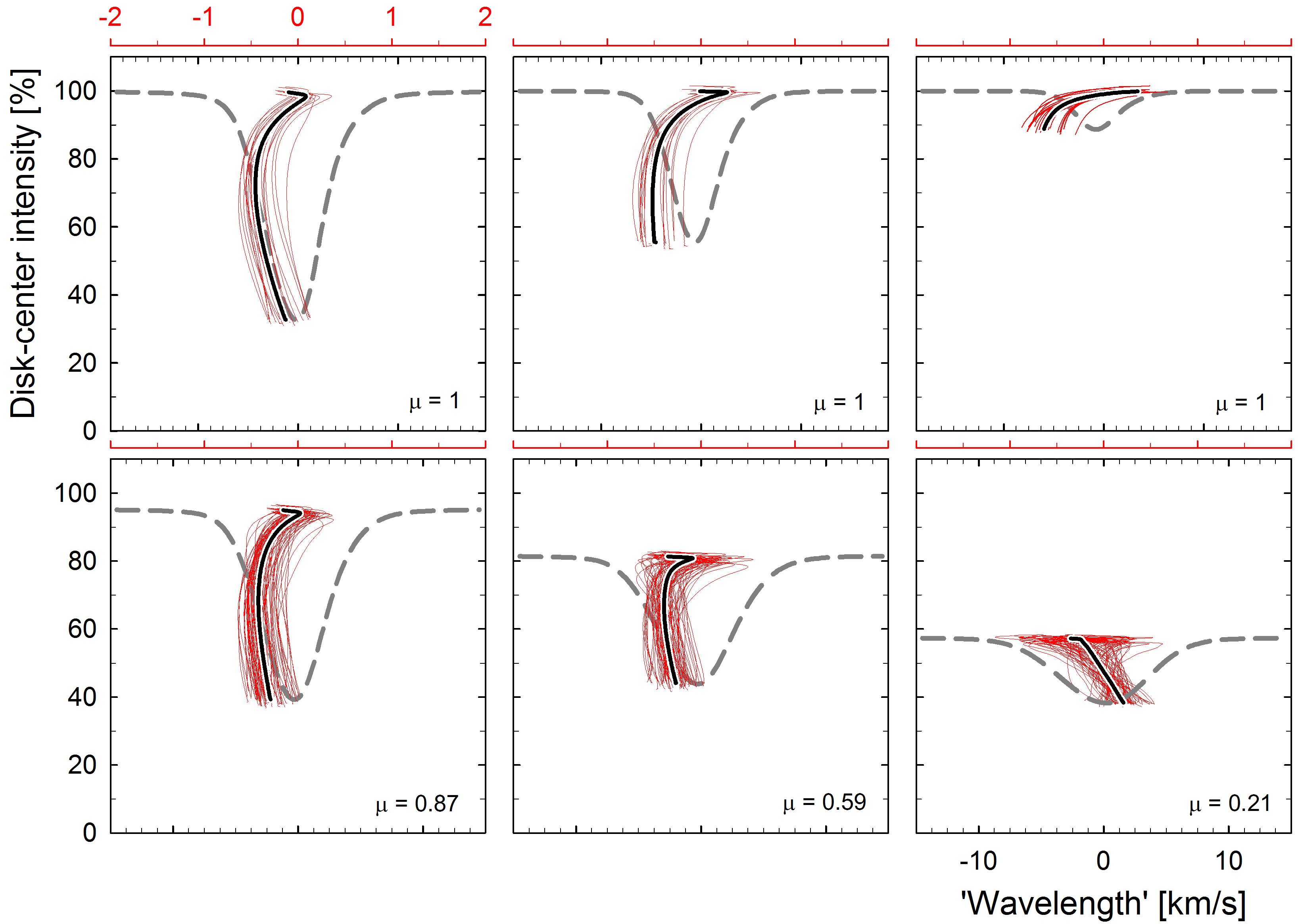}
\caption{Build-up of line asymmetries in synthetic Fe~I lines from the `F7~V' model.  Thin red lines (scales in km\,s$^{-1}$ at top) are bisectors of line profiles obtained as spatial averages for each of 20 temporal snapshots in the simulation, with global averages shown as bold black lines for the bisectors, and dashed for the line profiles (scales at bottom). Top: Stellar disk center ($\mu$ = 1), three different line strengths.  Bottom: Center to limb changes of the stronger line.  The bisector spread illustrates the time variability and also how extensive averaging that is required to reach a statistically stable signal.  Line {\it{shapes}} converge toward a stable mean (indicated by similar bisectors) sooner than their {\it{shifts}}.  The `blueward hook' of some bisectors near the continuum is an indicator of spatial inhomogeneities in line formation.}
  \label{fig:bisector_buildup}
\end{figure*}

\section{Theoretical signatures of 3-D atmospheres}

The above examples illustrate the need for and usefulness of a detailed understanding of both dynamic stellar atmospheres per se, their properties when used for exoplanet studies, and the spectral line formation in such atmospheres.  While many predictions from solar models have been verified against the observed solar surface,  corresponding tests for other stellar types are more challenging.  Detailed atmospheric properties can now be predicted for very different types of stars (even hypothetical stars that may lack real counterparts), while their testing is awkward for any star other than the Sun. 

However, the simulation of 3-D atmospheric hydrodynamics is only the first part of such analyses.  Any observational conclusions on, for example, stellar chemistry, surface velocity patterns, or exoplanet atmospheres require comparisons to synthetic spectra.  Such theoretical line profiles are obtained with the hydrodynamic simulation as a spatially and temporally varying set of model atmospheres, in which individual line profiles are computed for many spatial locations and at numerous timesteps, so that statistically stable values are reached.  Such calculations can be performed with different levels of detail and sophistication.  The most straightforward calculation is to use each surface element (or rather each vertical column of gas) of the simulation as a spatially local plane-parallel model atmosphere, computing line profiles at different inclination angles, assuming local thermodynamic equilibrium (LTE). 

Refinements may be introduced as departures from LTE, or even computing the full 3-D radiative transfer, where the spectrum from some location may depend on atmospheric properties not only along the line of sight.  For example, radiation emitted from hot vertical `walls' surrounding a cooler atmospheric depression, may illuminate the gas inside sideways and also scatter radiation in a direction toward the observer.  A huge number of atomic energy levels and transitions may potentially be relevant, but realistic calculations have to apply various simplifications to be manageable.  We now discuss what can practically be observed to guide and constrain such models. 

While 3-D simulations of atmospheric structure are carried out by several groups, only few have also calculated spectral line profiles across spatially resolved stellar disks, perhaps because their observability has not been appreciated.  On solar-type stars, with on the order of $\sim$10$^6$ granules across their surfaces, spectral line shapes on any significant fraction of the stellar disk are set by the averaged properties from very many granules and therefore are not sensitive to slightly different amounts of spatial smearing.  Similar to solar center-to-limb variations, the stellar surface properties (temperature contrasts, velocity fields, etc.) are reflected in the detailed line shapes, asymmetries, and wavelength shifts without requiring us to spatially resolve the stellar granulation pattern itself.  Thus, spectra from a modest number of center-to-limb positions may already provide unique information.

\subsection{Selected stellar models}

To explore possible observational signatures in different main-sequence stars, a group of 3-D simulations were produced with the CO\,$^5$BOLD code \citep{freytagetal12}, and synthetic spectral lines were computed from these simulations.

These models cover the range of solar-type dwarf stars with surface convection with temperatures T$_{\textrm{eff}}$ = 6730 ~K (approximate spectral type F3~V), 6250~K ($\sim$F7~V), 5900~K ($\sim$G0~V), 5700~K ($\sim$G2~V), and 3960~K ($\sim$K8~V).  Solar metallicity was applied throughout with model parameters given in Appendix Table A.1, where the designations within the CO\,$^5$BOLD grid are also specified, permitting the models to be uniquely identified. In addition, for comparison, one model of a giant star was also examined.  With more vigorous convection, its 3-D signatures in spectral lines are more pronounced.  However, the practical observability of its spatially resolved spectra using exoplanet transits appears not to be realistic since an exoplanet would cover only a very small fraction of its area (unless, perhaps, surrounded by an extensive and optically thick ring system), and it is not further discussed here.

\begin{figure*}
\sidecaption
  \includegraphics[width=12cm]{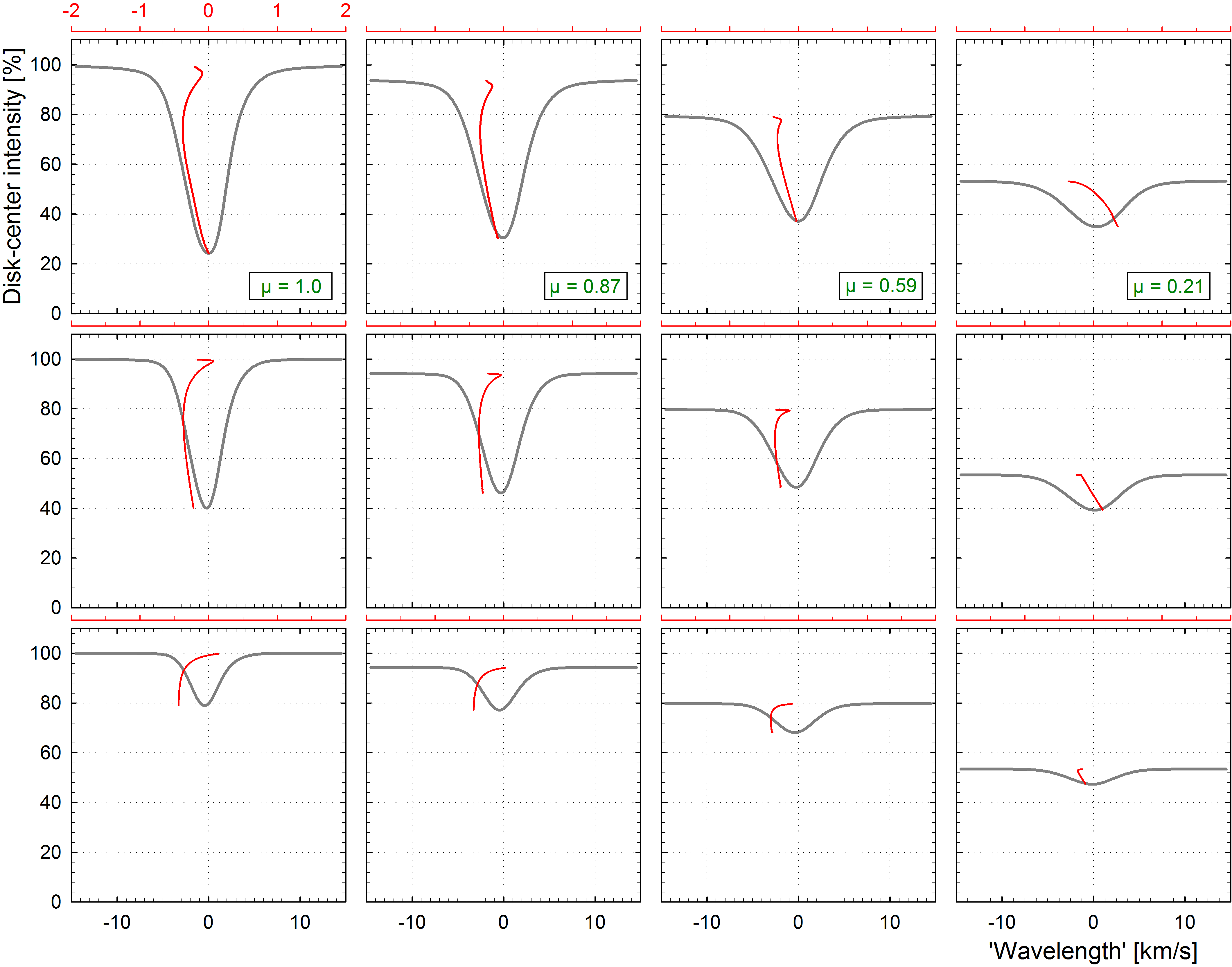}
     \caption{Synthetic line profiles (lower scale) and bisectors (upper scale) for the `G0~V' model, slightly hotter than the Sun.  Signatures of photospheric convection patterns are mapped onto a matrix of different behavior for lines of three different strengths at four different center-to-limb positions. }
     \label{fig:bisector_array}
\end{figure*}

\subsection{Synthetic spectral lines}

For each model, we computed synthetic Fe~I lines of five different oscillator strengths with lower excitation potential $\chi$ = 3 eV (sometimes also 1 and 5 eV), at $\lambda$ = 620 nm, for the different center-to-limb positions $\mu$ = cos\,$\theta$ = 1.00, 0.87, 0.59, 0.21, and four azimuth angles $\psi$ = 0, $\pi$/2, $\pi$, 3$\pi$/2 rad.  The spectral line synthesis was carried out in LTE, taking into account the full 3-D geometry of the flow, in particular including Doppler shifts caused by the velocity fields.  Thus, the radiative transfer equation was integrated along 13 different directions of the emerging radiation specified by these angle combinations in elevation and azimuth.  Each of these calculations was made for a grid of 47\,$\times$\,47 spatial points across the simulated stellar surface (downsampled from 140\,$\times$\,140 resolution elements in the actual hydrodynamic modeling), and for 19 or 20 temporally sampled snapshots during the extent of the simulation.  For each stellar model and each spectral line of any particular strength and excitation potential, this resulted in about half a million line profiles (each of 201 wavelength points covering $\pm$15~km\,s$^{-1}$), from which successive spatial and temporal averages were calculated.  For some hydrogen Balmer lines (H$\alpha$, H$\beta$, H$\gamma$), a broader wavelength range was covered, although an adequate interpretation of these strong lines would require a more complete modeling of the uppermost atmosphere.  Rotationally broadened profiles for full stellar disks were computed following \citet{ludwig07}.

\begin{figure*}
\sidecaption
  \includegraphics[width=12cm]{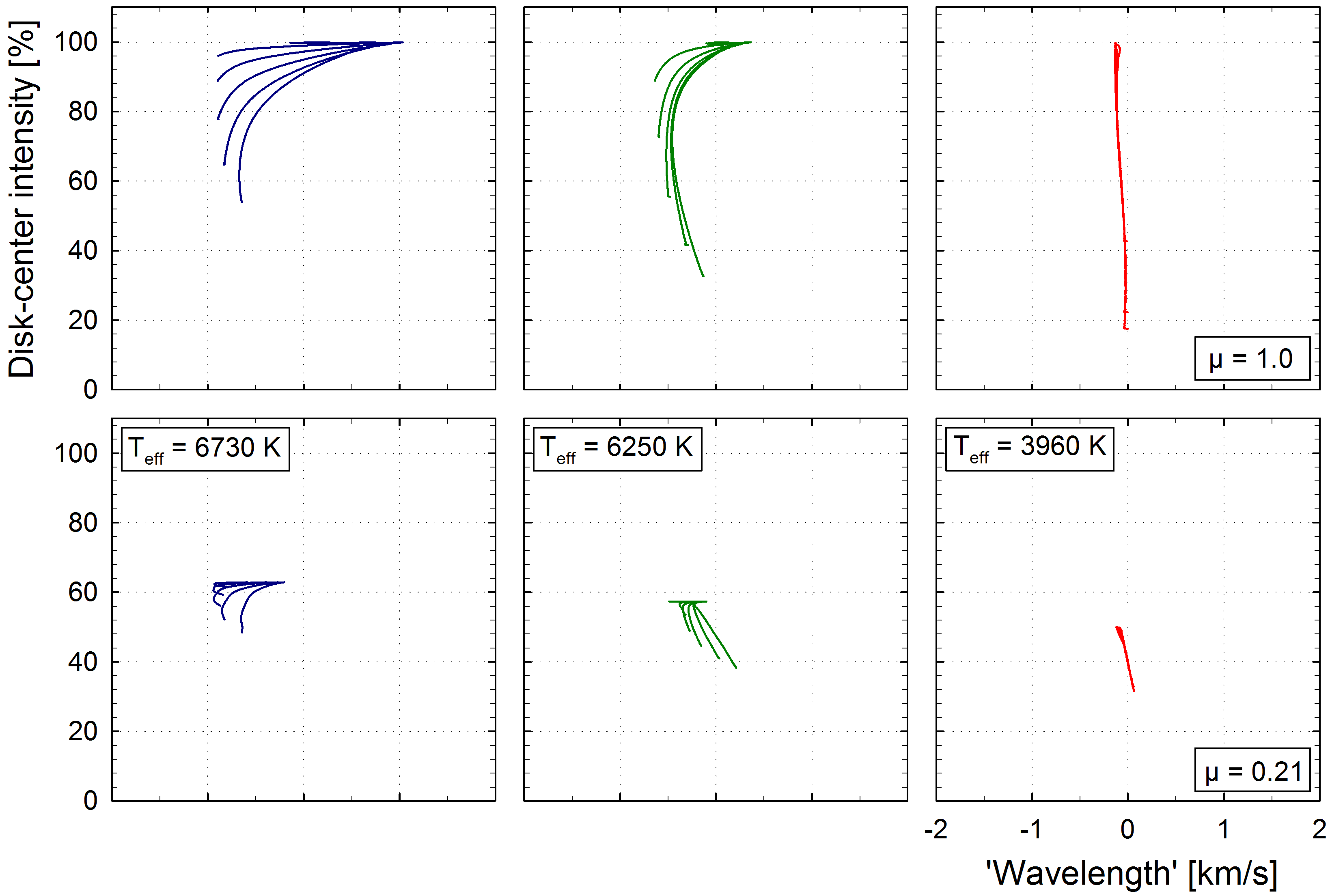}
     \caption{Synthetic bisectors show spectral line asymmetries for stars hotter and cooler than the Sun. Such bisectors reflect the vigor of surface convection in the line-forming layers and differ significantly between stars of different temperature.  Left to right: `F3~V', `F7~V', and `K8~V' are shown.  Top row shows bisectors for lines of five different strengths at stellar disk centers; bottom row near the limb at $\mu$ = 0.21. }
   \label{fig:bisectors_diff_stars} 
\end{figure*}

Figure \ref{fig:line_buildup} shows a typical build-up of representative spectral line profiles from successive spatial and temporal averages.  In the original simulation, individual line profiles spread over great intervals in wavelength and continuum intensity, since they originate in elements with quite different temperatures, moving with widely different radial velocities \citep[e.g.][]{asplundetal00}.   However, their spatial average begins to approach a representative profile, and stable profiles are obtained after averaging over many snapshots in time.  Since, in this particular star, the amplitude of convection increases with the depth of line formation, the 3-D signatures become more pronounced in weaker lines from deeper layers.  Figure \ref{fig:linewidths} shows another typical 3-D signature: spectral line broadening toward the stellar limb.  This results from horizontal motions that are typically greater than vertical motions, contributing more Doppler broadening when viewed along a line of sight that is closer to the limb.  In convective atmospheres, vertical velocities increase with height as the density decreases, until the release of radiation in the granular layers.  In the higher layers that are significant for photospheric spectral line formation, the vertical velocities rapidly decay, the gases turn over, and horizontal velocities dominate.  Examples of spatially resolved synthetic spectral lines are in \citet{dravinsnordlund90a}; for a discussion of the detailed mechanisms in model simulations, see \citet{nordlunddravins90} and \citet{nordlundetal09}. 

Smaller dependencies exist as functions of the excitation potential (Fig.~\ref{fig:excitationpot}) and ionization level.  Signatures also change between wavelength regions -- for a given temperature difference, the brightness contrast is greater at shorter wavelengths, correspondingly enhancing the statistical bias of brighter granules.  To evaluate even more subtle dependencies, the exact line formation physics has eventually to be accounted for precisely.

\subsection{Line asymmetries and convective wavelength shifts}

Photospheric line asymmetries and wavelength shifts may amount to only some percent of the full line widths and require some proper scale for meaningful display.  Figures \ref{fig:bisector_buildup}-\ref{fig:bisectors_diff_stars} show such line asymmetries and wavelength shifts in a bisector format.  In this amplified format, the spread in the bisectors between successive modeling timesteps becomes noticeable and indicates how long simulation sequences are required for statistical stability, which may differ among lines of different strength and at different stellar disk positions.  The linear extent of the simulation areas for our various models ranged between 5--16 Mm (Appendix A), which is rather smaller than a Jupiter-size exoplanet.

The relative bisector spread between modeling timesteps has a tendency to increase somewhat toward the stellar limb, especially for the hotter models.  This physical effect is caused by the stellar optical-depth surface being somewhat corrugated.  In averaging stellar areas near disk center, the line of sight reaches all spatial elements but near the limb, some elements sometimes become hidden behind local hills, adding another element of randomness. 

Bisectors of stronger lines sometimes have a sharply curved turn toward shorter wavelengths (`blueward hook') close to the continuum (Figs.\ \ref{fig:bisector_buildup}-\ref{fig:bisector_array}).  This can be traced to the extended Lorentzian-like wings of the stronger, saturated, and blueshifted line components.  Their contribution in one flank of the spatially averaged line also affects the intensity in the opposite flank in contrast to narrower Gaussian-like components, whose absorption disappears over a short wavelength distance.  First observed in the spectrum of Procyon, this mechanism was clarified by \citet{allendeetal02}, who predicted strong lines to show such a hook, but not weaker lines. The steep temperature gradients in the rising and blueshifted granules produce stronger absorption lines, which therefore tend to first saturate and develop Lorentzian damping wings in those spatial locations.  Such signatures in spatially averaged spectra thus reveal differences in line-formation conditions in different inhomogeneities across the stellar surface.  Figure \ref{fig:bisector_array} shows an example of an array of line profiles, bisectors, and convective wavelength shifts for lines of different strength and at different center-to-limb positions; such arrays begin to map atmospheric and line-formation properties throughout the entire photosphere. 

Figure \ref{fig:bisectors_diff_stars} shows bisector patterns for lines of five different oscillator strengths for stars of three different temperatures and at two different center-to-limb positions.  Besides striking differences in the line asymmetries and shifts, this plot -- with the intensity scales normalized to the disk centers -- also shows the successively more pronounced limb darkening in cooler stars.  The conspicuous line asymmetries in the hottest model reflects vigorous convection and `naked' granulation with full temperature fluctuations extending all the way up to the visible surface while the modest asymmetry in the coolest model results from its maximum temperature contrast being located slightly beneath the optically visible layers.  However, although the full temperature contrast in its granulation is hidden from direct view, the convective velocity fields extend up to the surface and determine line widths \citep{dravinsnordlund90a, nordlunddravins90, ramirezetal09}.  

Different surface geometries cause different changes of convective line shift across the star.  In stars with smooth photospheres, one expects the convective blueshift to decrease toward the limb, since the vertical convective velocities (correlated with granular brightness) then become perpendicular to the line of sight and the horizontal velocities contributing Doppler shifts appear symmetric.  Such type of behavior is found for the Sun and in stars of comparable temperature.  However, in the hottest model of Fig.~\ref{fig:bisectors_diff_stars}, there is an opposite tendency for the blueshift to {\it{increase}} from disk center toward the limb, caused by its corrugated photospheric surface (Sect.\ 2.2).

In any star, the precise amount of line asymmetry and shift depends on the atomic or molecular parameters.  High-excitation or ionized lines may be predominantly formed in the hottest elements, which are often the most rapidly rising and most
blueshifted;  thus, these lines show a more pronounced blueshift, while lines formed in high-lying layers of convective overshoot may experience an inverted correlation, instead resulting in a convective redshift.  Molecular lines predominantly form in the coolest elements (normally sinking and locally redshifted) and carry but little contributions from the hottest elements, where the molecules may have already dissociated, although very sophisticated modeling could also account for such time- and space-dependent molecular formation and dissociation.  Further details may depend on stellar chemical composition and the precise wavelength region.

\begin{figure}[H]
\centering
\includegraphics[width=\hsize]{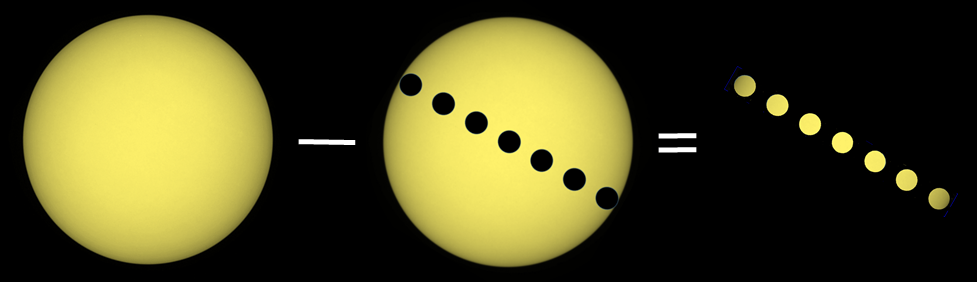}
\caption{Differences between stellar flux recorded outside and during exoplanet transit. Such differences yield a sequence of signatures from the temporarily hidden stellar surface segments.  The changing continuum flux is obtained from space-based photometry while the relative spectral changes are measured with ground-based spectrometers. }
\label{fig:stellar_subtracion}
\end{figure}

\section{Observable signatures}

Clearly, data of the types shown in Figs.~\ref{fig:line_buildup}--\ref{fig:bisectors_diff_stars} encode a great wealth of information on the stellar surface structure and, if sufficiently precise observational data can be obtained, strongly constrain possible models in terms of both atmospheric hydrodynamics and, not the least, of the ensuing spectral synthesis.  

A number of potentially feasible procedures to obtain spatially resolved stellar spectra were examined to understand which methods realistically could be already applied now or in the near future.  Perhaps, one could envision obtaining a diffraction-limited stellar image with an optical interferometer or a hypertelescope and then feed the light through some integral field spectrometer.  Another approach could be to monitor spectroscopically eclipsing binary stars to disentangle spectral contributions from each of the components.  Such methods could provide valuable information about stellar atmospheres (especially for giant stars) but it was concluded that, for high spatial resolution spectra across the surfaces of main-sequence stars, the cleanest and most direct method would be to use transiting exoplanets as scanning probes.    

The number of convective features (granules) varies between stars of different types, becoming fewer on giants and supergiants.  Solar-type stars have $\sim$10$^6$ granules across their surfaces and thus an exoplanet with projected area $\sim$1\% of the stellar disk covers $\sim$10$^4$ granules.  Simulations show that such a number is sufficient, even without further averaging over time, to produce characteristic and statistically stable spectral line profiles, even if the randomly changing number of granules causes its absolute wavelength to sway on a level of perhaps $\sim$10~m\,s$^{-1}$.  Simulations show, however, that the line {\it{shapes}} tend to become stable  with lesser averaging already (cf. Fig.\ \ref{fig:bisector_buildup}); this noise is thus primarily a swaying of the wavelength position of the line.  Assuming that the observed solar behavior is also representative for other solar-type stars, analogous effects should apply for stellar surface oscillations.  Although their local amplitudes reach $\sim$1\,km\,s$^{-1}$ over groups of many granules, the oscillations mainly cause the local line profiles to teeter in wavelength, while they average out adequately over larger areas. 

Thus, exoplanet areas are both small enough for highly selective spatial resolution and large enough to cover statistically stable line profiles in at least main-sequence solar-type stars.  Since the line profile changes across stellar disks are gradual, such profiles do not depend on the exact spatial resolution, i.e., the exact size of the planet,. This is the case as long as the planet is not very small nor subtends a significant fraction of a stellar diameter, although the latter would become an issue for red or brown dwarfs.

 \begin{figure*}
\sidecaption
  \includegraphics[width=12cm]{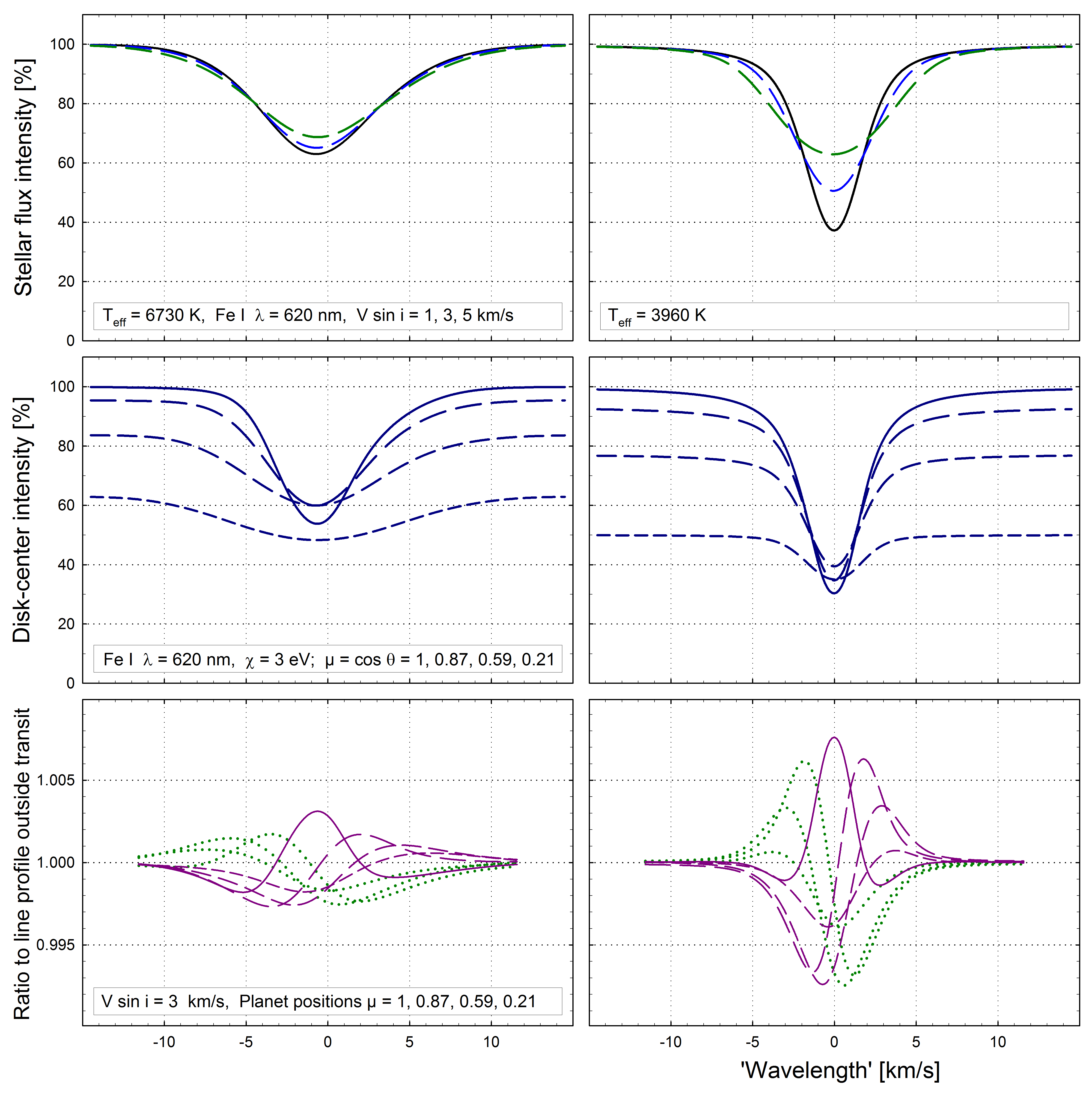}
     \caption{Observable signatures during the transit of an exoplanet covering 1.5\,\% of the surface in a hotter (`F3~V'; left) and a cooler star (`K8~V'; right).  Top row shows line profiles in integrated starlight for three rotational velocities $V$\,sin\,$i$ = 1, 3 and 5 km~s$^{-1}$.  Middle row shows spatially resolved line components at four different center-to-limb positions.  Such profiles can be disentangled from observations as in the bottom row: ratios of observed line profiles at each transit epoch to the line profile outside transit.  Red solid curves show these ratios at a planet position in front of stellar disk center.  Curves for positions from disk center toward the limb in a prograde planetary orbit are represented with successively shorter dashes.  Dotted green curves indicate the corresponding planet positions on the ingress side of the stellar disk; because of stellar line asymmetries, these do not exactly mirror those on the egress side. }
     \label{fig:hot_cool_dwarf_observable}
\end{figure*}

\subsection{Exoplanets and the observability of stellar surfaces}

During a transit, an exoplanet covers successive segments of the stellar disk and differential spectroscopy between epochs outside transit, and those during each transit phase, can provide spectra of each particular surface segment that was temporarily hidden behind the planet.  The method may appear straightforward in principle, but is observationally challenging since even Jupiter-size exoplanets cover only a tiny fraction of the stellar disk, $\sim$1\% of main-sequence stars.  If a desired signal to noise in the reconstructed spectrum were of order 100, for example, extracted from only $\sim$1\,\% of the total stellar signal, this would require an original S/N on the order of 10,000.  This is compounded by the need to observe during a limited time of the transit (or else averaging multiple transits).  This may appear daunting, but such spectral fidelity is not necessarily required for each individual spectral line.  Photospheric lines are numerous in cooler stars and offer the possibility of averaging even hundreds of these lines to recover the specific signatures from atmospheric structure which -- to a first approximation -- affects similar-type lines in an analogous manner, as seen in simulations (Fig.\ \ref{fig:bisector_array}) and observations alike \citep{dravins08}.

The desired observations during a transit are the absolute irradiance changes across spectrally resolved spectral lines.  However, directly such measurements are not practically feasible.  For ground-based observations, the small signal is compromised by varying atmospheric transmission and space measurements under adequate spectral resolution are unrealistic.  This is why some combination of data is required.  For spectral features, ratios between spectral line profiles at each transit epoch to the profile outside transit appear to be the least sensitive to systematic and random noise.  Such ratios can be obtained with wavelength-stable, high-resolution, ground-based spectrometers.  Their absolute continuum-flux levels reflecting irradiance changes can then be calibrated with precise space-based photometry with lower spectral resolution.  Since the transits are repetitive, these measurements need not be from the same transit and also data from several transits may be combined or averaged.

To disentangle the spectrum from behind the planet, further data are needed, in particular the effective area $A_{\textrm{eff}}$ of the planet.  At each transit position, the amount of obscured flux is proportional to the product of the projected geometric area of the planet with the stellar continuum brightness in the relevant wavelength region.  This product is directly obtained from the photometric transit curve but to deduce the geometric area of the planet requires us to understand the projected path of the planet and its location on the stellar disk.  These parameters are obtained from radial-velocity measurements of the Rossiter-McLaughlin effect, which is discussed below.  

The temporarily hidden spectral flux $S$ at any given transit phase thus equals that from the full stellar disk (outside transit) minus that from the non-obscured portions, its normalization in intensity set by the effective planet area: $S_{\textrm{hidden}}$ = ($S_{\textrm{outside-transit}}$ – $S_{\textrm{in-transit}})/A_{\textrm{eff}}$.  Besides the absolute flux levels, a precise wavelength calibration is also required.  An error analysis shows that precisions on the order of $\sim$10~m\,s$^{-1}$ in radial velocity are necessary for the later retrieval of spectral line profiles.  Those are obtained as tiny differences between the out-of-transit and transit profiles and even a small error in the wavelength displacement of either may cause a significant deviation in the reconstruction.  This implies that the spectral wavelength scales must be corrected not only to heliocentric values that compensate the motion of the observer relative to the solar system barycenter, but to {\it{astrocentric}} values, also accounting for the motion of the stellar center-of-mass induced by the orbital motion of its exoplanet, which thus also must be precisely known.

 \begin{figure*}
\sidecaption
  \includegraphics[width=12cm]{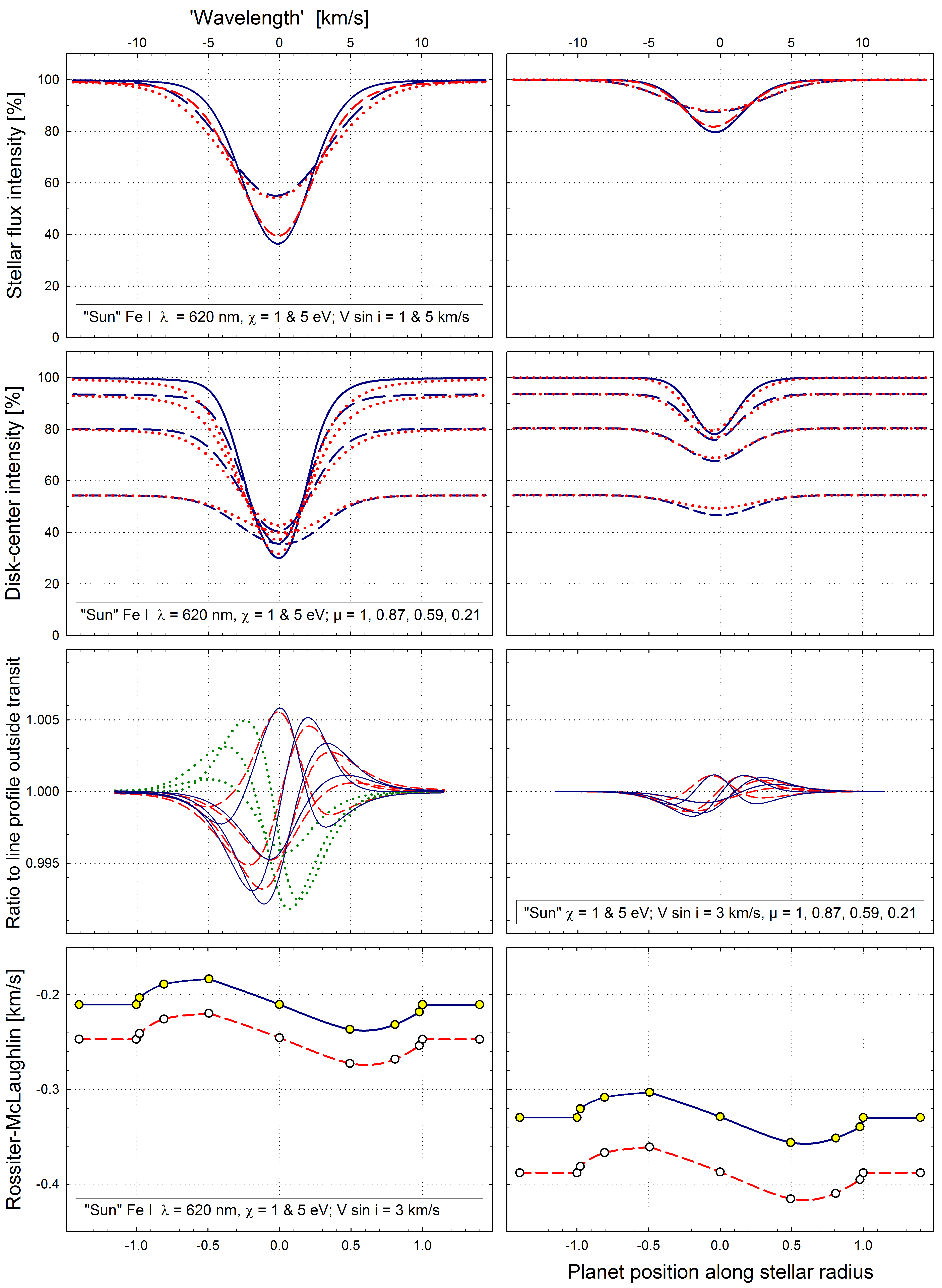}
     \caption{Observable signatures during a transit of an exoplanet covering 1.5\,\% of the stellar surface across a solar-temperature star for differently strong spectral lines, different excitation potentials, and different velocities of stellar rotation.  Top row shows synthetic Fe~I line profiles in integrated starlight for rotational velocities $V$\,sin\,$i$ = 1, and 5 km\,s$^{-1}$, and for excitation potentials $\chi$ = 1 eV (blue dashed) and 5 eV (red dotted).  The second row shows the spatially resolved lines at different center-to-limb positions and the third row shows the observable ratios between line profiles during and outside transit, as in Fig.~\ref{fig:hot_cool_dwarf_observable}.  The bottom row shows the Rossiter-McLaughlin signal, i.e., the apparent radial velocity of the star relative to its true center-of-mass motion, obtained as a Gaussian fit to the line profile in integrated starlight (however, neglecting gravitational redshift).  The displacements between the curves (1 eV -- solid blue; 5 eV -- dashed red) reflect the dependence of convective blueshift on excitation potential.}
     \label{fig:sun_observable}
\end{figure*}

\subsection{Simulated line changes during exoplanet transits}

It does not seem realistic (at least from the ground) to precisely measure absolute changes in stellar spectral line irradiance during a transit. Instead, some relative ratios have to be measured.   An examination of various options suggests that ratios between spectral line profiles during any transit epoch to those outside transit appear to be the least sensitive to systematic and random noise.   Although  the continuum flux levels also need to be calibrated; this can be carried out with space-based photometry with lower spectral resolution. 

Figure \ref{fig:hot_cool_dwarf_observable} shows expected line profile ratios during exoplanet transits across the equator for both hotter (T$_{\textrm{eff}}$ = 6730 K), and cooler (T$_{\textrm{eff}}$ = 3960 K) main-sequence stars of solar metallicity, rotating with $V$ = 3 km\,s$^{-1}$.  Red curves show the ratios of the instantaneous line profiles relative to those outside transit.  These simulation sequences from a CO\,$^5$BOLD model are for an Fe~I line ($\lambda$ = 620 nm, $\chi$ = 3 eV) during the transit by an exoplanet in a prograde orbit, covering 1.5\,\% of the stellar surface.  For moderate changes of the planetary area, the signal changes in a linear fashion.  The cooler model produces narrower lines and greater amplitudes in their ratios.  However, even so, the amplitudes do not exceed $\sim$0.5\,\%, requiring noise levels below $\sim$0.1\% for detection.

Figure \ref{fig:sun_observable} shows the corresponding signatures for the solar model (T$_{\textrm{eff}}$ = 5700 K), but now separately for low- and high-excitation lines ($\chi$ =1 and 5 eV) for two rotational velocities (1 and 5~km\,s$^{-1}$) and two different line strengths.  The differential signatures between low- and high-excitation lines are an order of magnitude smaller but even those should be observable, at least in principle.

\subsection{Observational requirements}

While observational data of exceptionally high quality are required, obtaining such data is  feasible.  Data for actual stars will be evaluated starting with Paper II \citep{dravinsetal17} but here, we already give an example of spectroscopic observations from which observable signatures can start to be extracted.

\begin{figure}[H]
\centering
\includegraphics[width=8 cm]{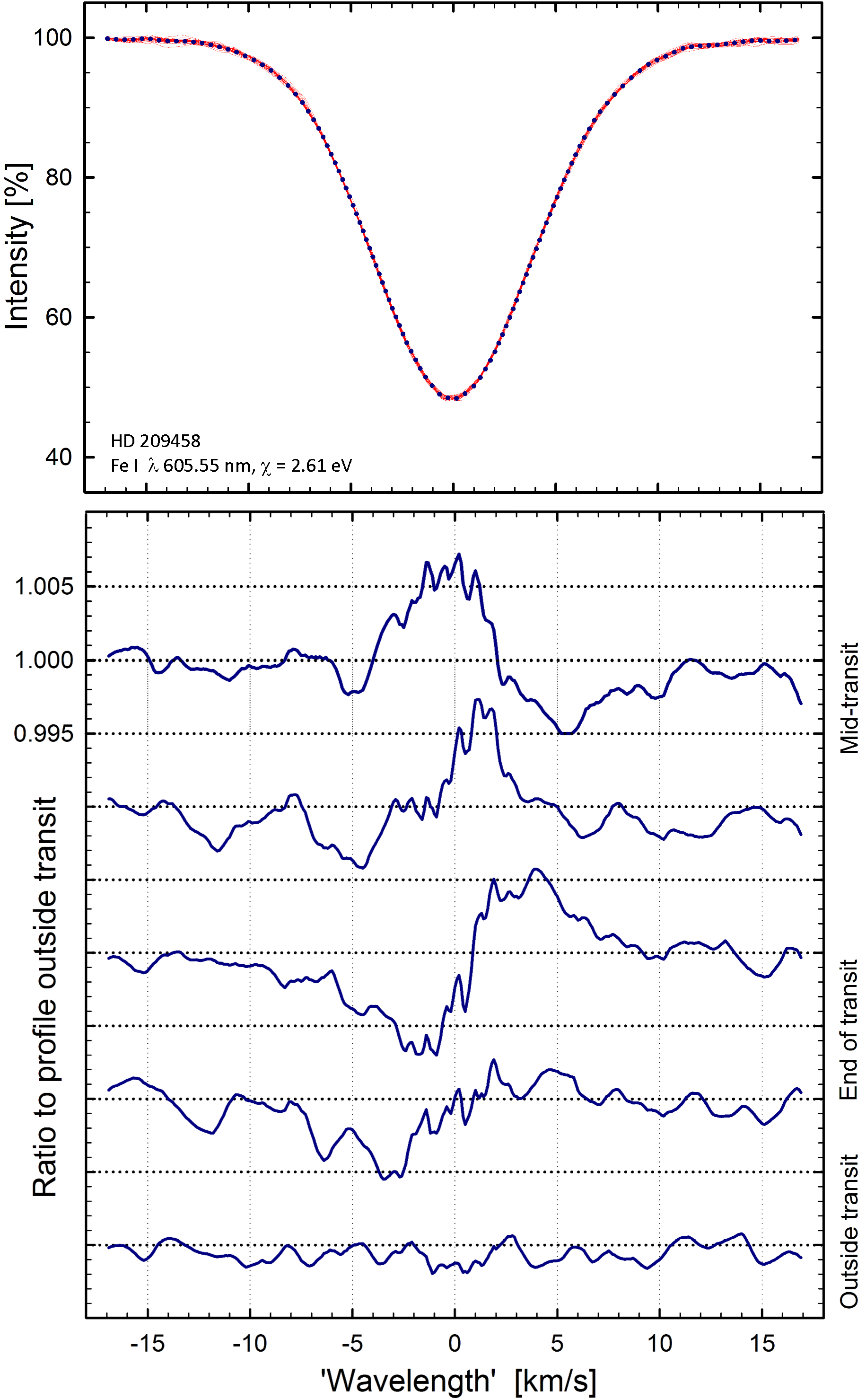}
\caption{ Top: Observed line profiles for one photospheric Fe~I line in the planet-hosting star HD~209458.  The thin red curves are observed profiles from various epochs during the exoplanet transit; the dotted curve is the reference profile from outside transit.  Bottom: By averaging data from 11 different Fe~I lines, the photometric signal-to-noise ratio approaches what is required to resolve the gradual line profile changes during transit.  The line profile ratios are in the same format as the theoretical curves in Figs.\ \ref{fig:hot_cool_dwarf_observable} and  \ref{fig:sun_observable}. }
\label{fig:sample_observation}
\end{figure}

Figure \ref{fig:sample_observation} shows one essentially unblended relatively strong Fe~I line in the planet-hosting star HD~209458 (G0~V) recorded with the VLT-UVES spectrometer at ESO with a spectral resolution $\lambda/\Delta\lambda\sim$80,000.  Although largely overlapping on the plotted scale, the figure shows profiles from numerous successive exposures during an exoplanet transit, together with a reference profile obtained as an average of exposures outside transit.  These data are representative for current best individual stellar spectra with photometric signal-to-noise ratios for individual exposures around $\sim$500 or somewhat higher.  However, even such data are not quite adequate, but by stacking 11 Fe I lines of comparable strength, an averaged profile is obtained with a nominal S/N approaching $\sim$2,000.  The ratios of this averaged Fe~I profile at each transit epoch to that outside transit, now begin to show signatures of the type theoretically expected in Figs.\ \ref{fig:hot_cool_dwarf_observable} and \ref{fig:sun_observable}.  This illustrates the minimum requirements but obviously even lower noise data are desired.  Such data can be obtained by averaging over a greater number of spectral lines, over multiple transits, or simply with better original measurements.

\subsection{Rossiter-McLaughlin effect}

During the transit of an exoplanet across a rotating star, it selectively hides different portions of the stellar surface where the local rotational velocity vector is toward or away from the observer.  Thus, part of the blue- or redshifted photons are removed from the integrated starlight, whose averaged wavelength then appears slightly red- or blueshifted.  The geometry of the projected path of the exoplanet across the star (including whether it moves in a prograde or retrograde orbit) must be known to fully interpret any observed line profile variations.  That can be determined through this Rossiter-McLaughlin effect; e.g., \citet{perryman11} or \citet{albrecht12}.  The effect has been measured for numerous exoplanets, while the specific issue on the influence of convective wavelength shifts and stellar line profile changes across stellar disks has been evaluated by \citet{ceglaetal16}, \citet{reinersetal16a}, \citet{shporerbrown11}, and \citet{yanetal15a, yanetal15b, yanetal17}. 

The Rossiter-McLaughlin signal, expressed as an apparent radial velocity, is not a uniquely defined quantity since the full spectral line profile at any one transit epoch is then reduced to one single number.  Since all line profiles are somewhat asymmetric, the exact value of a line profile depends on exactly how the line is approximated or fitted. 

\begin{figure}[H]
\centering
\includegraphics[width=8 cm]{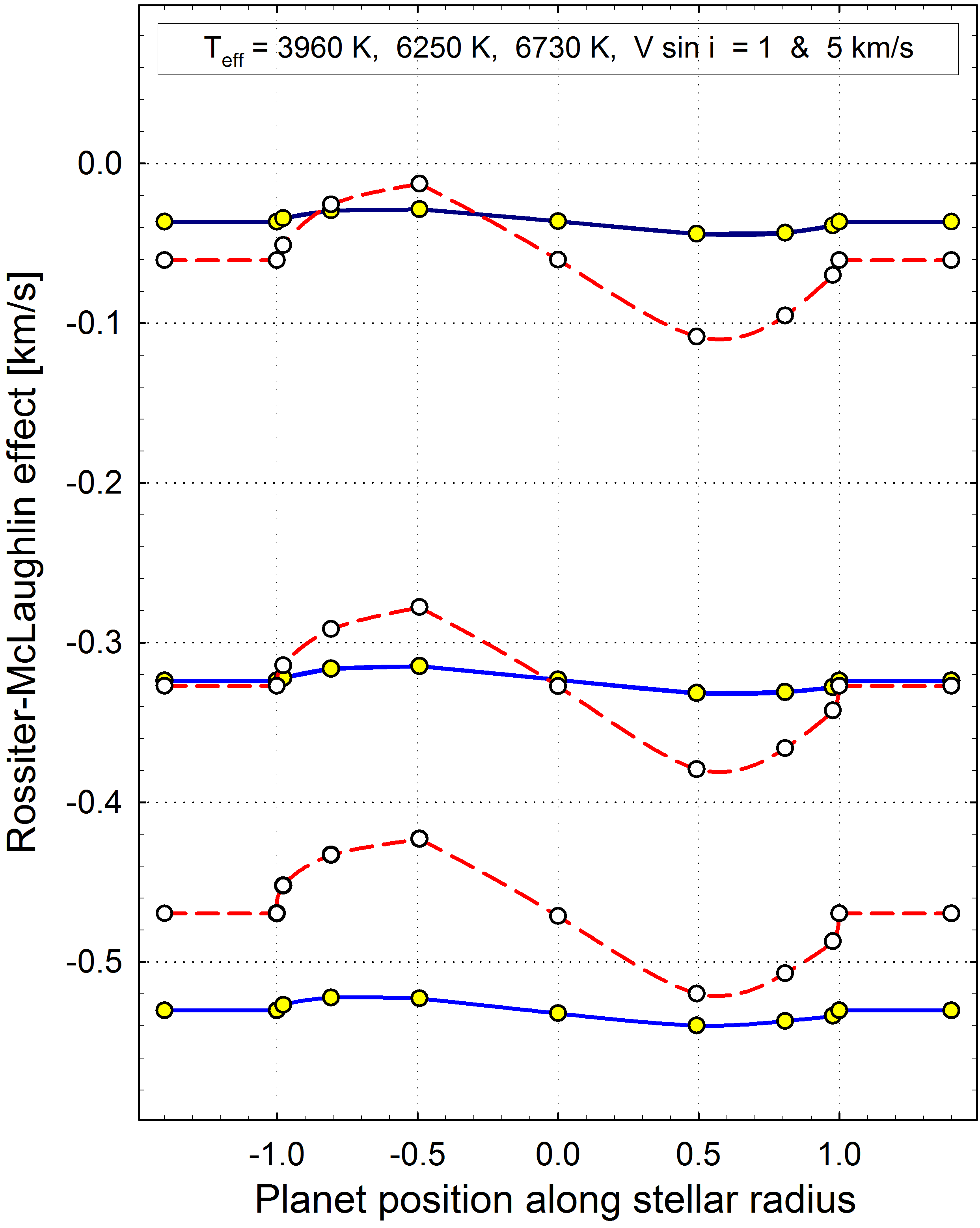}
\caption{Apparent radial velocity of rotating stars changes during planetary transits owing to the Rossiter-McLaughlin effect, here modeled for stars of three different temperatures (`K8~V', `F7~V', and `F3~V') for the rotational velocities $V$\,sin\,$i$ = 1 and 5 km\,s$^{-1}$ (solid blue and dashed red, respectively).  The planet was assumed to cover 1.5\,\% of the stellar disk area with the transit along the stellar equator.  The signal was obtained as a Gaussian fit to the line profile in integrated starlight.  The wavelength values reflect different amounts of convective blueshift and are also affected by rotational line broadening, while the amplitudes are a function of rotational velocity.  Gravitational redshifts are neglected. }
\label{fig:r-m_effect}
\end{figure}

 This value also depends on line broadening caused by either stellar rotation or finite spectroscopic resolution; even a symmetric instrumental profile, convolving an asymmetric line, causes a profile with a different asymmetry.  In  Figs.~\ref{fig:sun_observable} and \ref{fig:r-m_effect}, synthetic line profiles were fitted with five-parameter modified Gaussian functions of the type $y_0 + a \cdot \exp[-0.5\cdot(|x-x_0|/b)^c]$: the plotted values show the fitted central wavelengths.

The main difference between the curves for different excitation potentials in the bottom row of Fig.~\ref{fig:sun_observable} and for different stars in Fig.~\ref{fig:r-m_effect} lies in their absolute wavelength positions, i.e., different values of their respective convective wavelength shifts.  Even if the profile {\it{shapes}} for lines of different excitation potentials or ionization levels might be closely similar, they may be segregated through different {\it{shifts}}.  An observational advantage is that the Rossiter-McLauglin signature is easier to measure than line profiles and differences between different stars, such as in Fig.~\ref{fig:r-m_effect}, are significant.  To identify such effects within any one star, however, requires accurate values of the laboratory wavelengths of the respective lines (or their average for groups of lines), and when comparing different stars, their relative physical radial motions must be precisely known.  A minor effect is that the curves are not exactly symmetric with respect to the stellar disk center, which is an effect of the convective blueshift; cf.\ \citet{dravinsetal15}.   Possibly, for the precise calculation of line shifts in synthetic profiles, their atmospheric formation height also needs to be well understood since differences in the gravitational redshift between different atmospheric layers might be not entirely negligible \citep{ceglaetal12, lindegrendravins03}. 

Studies of such absolute wavelength shifts are more straightforward for the Sun, where the relative Sun-Earth motion is well known from planetary system dynamics and does not rely on wavelength measurements of spectral lines.  Options for other stars include astrometric determination of absolute radial motion of stars in moving clusters \citep{madsenetal02} or spectroscopic determination of relative motions within open and undispersed clusters, whose stars share a common velocity \citep{pasquinietal11}.

\section{Observational outlook}

Spatially resolved stellar spectra do not suffer any rotational broadening since they are not averaged across the full disk and, in principle, reveal the intrinsic line shapes, merely Doppler-shifted by the rotational velocity vector at each particular stellar disk location.  Spectral line synthesis from hydrodynamic simulations predicts a rich diversity of phenomena in the photospheric line profile shapes, asymmetries, and shifts across stellar disks.  If even only a fraction of those can be observed, this could considerably constrain such models.  Further, if a transiting planet would happen to cross a starspot, for example, even spatially resolved spectra (with their magnetic signatures) of such stellar surface features could become attainable, given sufficient spectral fidelity, and thus contribute to the magnetohydrodynamics in stellar atmospheres as well.  Once sufficiently bright host stars with sufficiently large transiting planets are found or sufficiently large telescopes are available, such studies could be extended to stars with special properties, perhaps rapidly rotating stars, those with strong stellar winds, and other classes.  However, the use of transiting exoplanets as probes appears practical only for main-sequence dwarf stars, where even a planet of Jupiter size subtends not much more than $\sim$1$\,\%$ of the stellar disk.  

Starting with Paper II of this project \citep{dravinsetal17}, we will demonstrate how the retrieval of spatially resolved stellar spectra is practically feasible with current facilities already, at least for a few well-observed stellar targets.

\begin{acknowledgements}
{HGL acknowledges financial support by the Sonderforschungsbereich SFB881 `The Milky Way System' (subproject A4) of the German Research Foundation (DFG).  The work by DD was performed in part at the Aspen Center for Physics, which is supported by National Science Foundation grant PHY-1066293.  DD also acknowledges stimulating stays as a Scientific Visitor at the European Southern Observatory in Santiago de Chile.  We thank the referee for precisely pointing out some sections in need of elucidation. }

\end{acknowledgements}

\begin{appendix}

\section{Hydrodynamic models}

The models used for the hydrodynamic simulations and the ensuing spectral line synthesis are characterized by stellar parameters, such as temperature and surface gravity, and by their spatial extent and computational step sizes.  In the text these models were mainly referred to by their approximate spectral type.  The listing of model identifiers in Table A.1 enables their unique identification or comparison with other work based on the CO\,$^5$BOLD grid \citep{freytagetal12}. 

The number of sampling points for each spectral line profile is the product of horizontal spatial resolution points in $x$ and $y$, the number of temporal snapshots extracted during the simulation sequence, and the number of different directions (combined elevation and azimuth angles) for the emerging radiation.  The horizontal ($hsize$) and vertical ($vsize$) extents of the hydrodynamic simulation volume are here given in cm.

\begin{table*}
\caption{CO\,$^5$BOLD model identifiers}             
\label{table:1}      
\centering          
\begin{tabular}{c c c c c c c}
\hline\hline       
T$_{\textrm{eff}}$ [K] & log~$\varg$ [cgs] & Approx.sp.\ & Model & Spectral points & boxsize:$hsize$ [cm] & boxsize:$vsize$ [cm] \\ 
\hline                    
6730 & 4.25 & `F3~V' & d3t68g43mm00n01 & 47 $\times$ 47 $\times$ 20 $\times$ 13 & 1.64459e+09 & 2.41691e+09 \\  
6250 & 4.5 & `F7~V' & d3t63g45mm00n01 & 47 $\times$ 47 $\times$ 20 $\times$ 13 &  7.00000e+08 & 3.94843e+08 \\
5900 & 4.5 & `G0~V' & d3t59g45mm00n01 & 47 $\times$ 47 $\times$ 19 $\times$ 13 &  6.02000e+08  & 3.77875e+08 \\
5700 & 4.4 & `G2~V' & d3gt57g44n58 & 47 $\times$ 47 $\times$ 19 $\times$ 13 & 5.60000e+08 &  2.25378e+08 \\
3960 & 4.5 & `K8~V' & d3t40g45mm00n01 & 47 $\times$ 47 $\times$ 19 $\times$ 13 & 4.73587e+08 & 1.22689e+08 \\
\hline                  
\end{tabular}
\end{table*}

\end{appendix}

\end{document}